\documentclass[twocolumn]{aa}

\usepackage{graphicx}
\usepackage{color,txfonts}
\usepackage[colorlinks=true]{hyperref}
\hypersetup{citecolor= blue, linkcolor=red, urlcolor=blue}
\usepackage{microtype}

\newcommand{\Hazel}{{\sc Hazel}}

\newcommand{\MultiNest}{{\sc MultiNest}}

\newcommand{\Helix}{{\sc HeLIx}}
\newcommand{\kms}{\,km\,s$^{-1}$}

\begin{document}

   \title{Spectropolarimetric analysis of an active region filament}
   \subtitle{II. Evidence of the limitations of a single component model}

   \author{D\'iaz Baso, C. J.
          \inst{1,2,3}
          \and
          Mart\'{\i}nez Gonz\'alez, M. J.
          \inst{1,2}
          \and
          Asensio Ramos,  A.
          \inst{1,2}
          }

   \institute{Instituto de Astrof\'isica de Canarias, C/V\'{\i}a L\'actea s/n, E-38205 La Laguna, Tenerife, Spain
   \and
   Departamento de Astrof\'{\i}­sica, Universidad de La Laguna, E-38206 La Laguna, Tenerife, Spain
   \and
   Institute for Solar Physics, Dept. of Astronomy, Stockholm University, AlbaNova University Centre, SE-10691 Stockholm Sweden
             }

   \date{Received December 06, 2018; accepted April 21, 2019}

   \authorrunning{D\'iaz Baso et al.}
   \titlerunning{Magnetic field of an active region filament}

  \abstract
   {}
   {Our aim is to demonstrate the limitations of using a single component model to study the magnetic field of an active region filament. For that we have analyzed the polarimetric signals of the \ion{He}{i}~10830\,\AA\ multiplet acquired with the infrared spectrograph GRIS of the GREGOR telescope (Tenerife, Spain).}
   {After a first analysis of the general properties of the filament using \Hazel\ under the assumption of a single component model atmosphere, in this second part we focus our attention on the observed Stokes profiles and the signatures which cannot be explained with this model.}
   {We have found an optically thick filament where the blue and the red components have the same sign in the linear polarization as an indication of radiative transfer effects. Moreover, the circular polarization signals inside the filament show the presence of strong magnetic field gradients. We have also shown that even a filament with such high absorption still shows signatures of the circular polarization generated by the magnetic field below the filament. The reason is that the absorption of the spectral line decays very quickly towards the wings, just where the circular polarization has a larger amplitude. In order to separate both contributions, we explore the possibility of a two component model but the inference becomes impossible to overcome as a high number of solutions is compatible with the observations.}
   {}


   \maketitle

\section{Introduction}\label{sec:intro}

After some decades of challenging observations of solar filaments and prominences and the development of the theory that explains how polarization signals are generated and modified due to the magnetic field \citep[see][and references therein]{Bommier1981,Landi1982,bommier1989,TrujilloBueno2002,TB2007,CASINI2009}, still we do not know exactly the details of their magnetic topology \citep{Labrosse2010,mackay2010}. Numerous models and simulations  have allowed a better understanding of the formation process and how this plasma is supported against gravity  \citep{Kippenhahn1957,Kuperus1974,Antiochos1999,Xia2012,Keppens2014}, but the most reliable way to obtain quantitative information on its magnetic field and verify the veracity of the current models is through the analysis of spectropolarimetric observations.

Solar filaments can be observed, among other spectral lines, in the \ion{He}{i}~10830\,\AA\ multiplet as absorption features. This spectral line has been widely used as a chromospheric diagnostic tool because of its sensitivity to the Hanle and Zeeman effects and the release of fast and robust inversion codes, such as \Hazel\ \citep{asensio2008} and \Helix$^+$ \citep{lagg2004,lagg2009}. This spectral line is commonly optically thin in the solar spectrum and enough opacity is only reached around specific patches over the solar disk. This multiplet also suffers from several problems. First, the involved polarization signals are weak and only at a level of a few $10^{-3}$~$I_c$ in the better cases\footnote{
This value corresponds to the typical linear polarization measured in prominences/filaments but also in other solar regions \citep{schad2013}. Even in AR filaments with high horizontal magnetic fields a maximum of $\sim3.10^{-3}I_c$ is reached \citep{kuckein2012}. On the contrary, Stokes $V$ can reach higher values of the order of $10^{-2}I_c$ in very magnetized regions.
}. Second, the presence of magnetic ambiguities, leads to different magnetic topologies that generate similar polarimetric signals. These ambiguities can be solved if the configuration is expected to be simple such as in sunspots \citep{schad2013,joshi2016} where the expected magnetic field is assumed to be nearly radial from the center of the sunspot aligned with the filaments of the inner penumbra. However, filaments and prominences are very difficult to disambiguate. A good example is shown by \cite{marian2015} who imposed some additional physical constraints on the stability of the structure to reject some of the inferred configurations, or \cite{orozco2014} whose selection was based on the chirality of the filament.

New recent studies have shown the importance of considering more complex models to understand the physical mechanisms behind the formation of the \ion{He}{i}~10830\,\AA\ multiplet. There are observational evidences, for example, of the generation of atomic orientation by the Doppler shifted illumination coming from the underlying magnetized photosphere, which explains the existence of extremely asymmetric Stokes $V$ profiles \citep{MartinezGonzalez2012}. Other studies point out, from numerical experiments, the problems of assuming a 1D slab model to infer the magnetic field of complex structures \citep{milic2016}. They emphasize the multidimensional effects often neglected in optically thick structures, also noted by others  from the ratio between the two components of the helium triplet \citep{LopezAriste2002}.

Concerning filaments on-disk, \cite{diaz2016} showed that the polarization signals measured in active region filaments above magnetized regions \citep{kuckein2009,xu2012} could be subject to biases. They pointed out that the observed signals could be explained with a two component atmospheric model simulating the filament and the chromospheric region below. This is a very natural way of explaining the observations without relying on other mechanisms like the presence of horizontal illumination in the slab \citep{TB2007} or unresolved magnetic fields \citep{CASINI2009}, which have been proposed to strongly modify the polarization signals.

We studied, in the first part of this series \citep[][hereafter ``Paper~I'']{DiazBaso2019A} high quality polarimetric observations of an active region filament located above granulation. Even being less contaminated by the underlying magnetized areas (a sunspot), a single  component model showed some unrealistic and inconsistent results in the magnetic field vector inference. The Stokes $V$ map of \ion{He}{i} does not show any clear signature of the presence of the filament while it is very similar to the circular polarization map of the photosphere. After the one component inference, the local azimuth map follows the same pattern observed in the circular polarization map. This means that Stokes $V$ is conditioning the inferred magnetic field vector producing unrealistic results in the spine of the filament (strong shears in the azimuth) that suggests the Stokes $V$ could come fundamentally from a different place than the linear polarization. In this second part we describe some clear hints of the necessity of more complex models to explain the observed profiles. Simple one component models are not able to fully reproduce them and more importantly, can lead to biased or unphysical results.

\section{Observations}\label{sec:observation}

\begin{figure}[ht!]
\centering
\includegraphics[width=\linewidth]{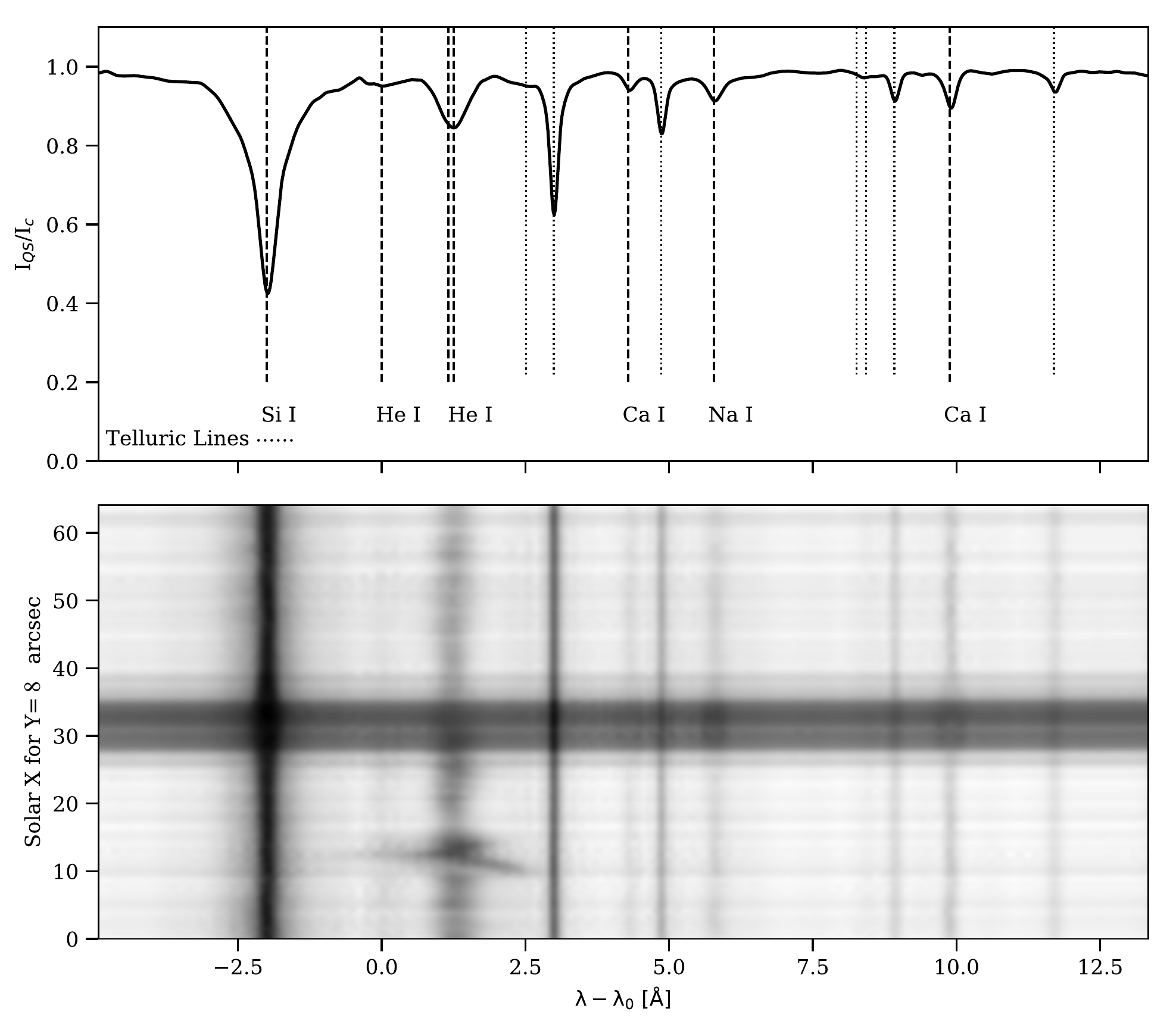}
\caption{Upper panel shows the median intensity spectrum of the quiet Sun normalized to
the quiet Sun continuum. The lower panel shows the spectra along the slit for the third scanning position. The reference wavelength \textrm{$\lambda_{0}=10829.0911$}\,\AA\
is the central wavelength of the blue component of the \ion{He}{i} 10830\,\AA\ multiplet. Close to it, the two red components of the same multiplet are blended in the same profile.}
\label{fig:RANGO1}
\end{figure}

The observations were carried out on 17th of June 2014 using GRIS \citep[GREGOR Infrared Spectrograph,][]{gris2012} at the GREGOR telescope \citep{gregor2012}. We  recorded the spectropolarimetric measurements of an active region filament located close to the disc center ($\mu=0.92$). The observed filament was situated close to a sunspot in the active region NOAA 12087. The data acquisition and reduction process were described in detail in Paper~I. An example of the observed spectra is displayed in Fig.~\ref{fig:RANGO1}. In the upper panel we see the corrected Stokes $I$ spectra obtained as the average profile of a quiet Sun area. The lower panel shows a slit-spectra chosen from the third scan at $\mathrm{Y=8}$\,\arcsec. In the lower part close to $\mathrm{X=12}$\,\arcsec\ we can see the effect of the highly dynamic chromosphere on the \ion{He}{i} triplet at 10830\,\AA. These kind of profiles are found in the border of the filament with a clear second component strongly red-shifted with velocities around 30\kms. These pixels can be easily explained as material falling down from the filament, where the magnetic field is more vertical and cannot keep the material against gravity. This case is an obvious example of the necessity of an additional atmospheric component as they are clearly shifted in wavelength. This scenario has also been reported by other studies \citep{sasso2011,schad2016}, and usually happens in very localized regions of the field of view (FOV). In this study we point out many evidences of the presence of multi-component atmospheres in a large fraction of the FOV.

Therefore, in the following sections we analyze in detail the observed profiles, showing how a single component model is not able to reproduce certain aspects in the intensity, linear polarization and circular polarization profiles. For this purpose we make use of the third scan because the filament has the highest absorption and the features are more noticeable,  but similar conclusions can also be drawn from the other scans.

\section{Analysis of Stokes $I$}\label{subsec:10830stokeI}

Numerous studies of solar prominences have inferred the physical parameters of the plasma, and in particular their temperature by analyzing the shape and intensity of spectral lines when the plasma is optically thin \citep[see review of][and references therein]{Parenti2014}. However, we demonstrate that our observations show signs of the presence of optically thick plasma \citep{LopezAriste2002}. In the following, we show two consequences of the large absorption in the filament: i) the differential saturation effect between the red and the blue components of the triplet and, ii) misfits when using \Hazel\ which can also be explained by radiative transfer effects.

\begin{figure}[!ht]
\centering
\includegraphics[width=\linewidth]{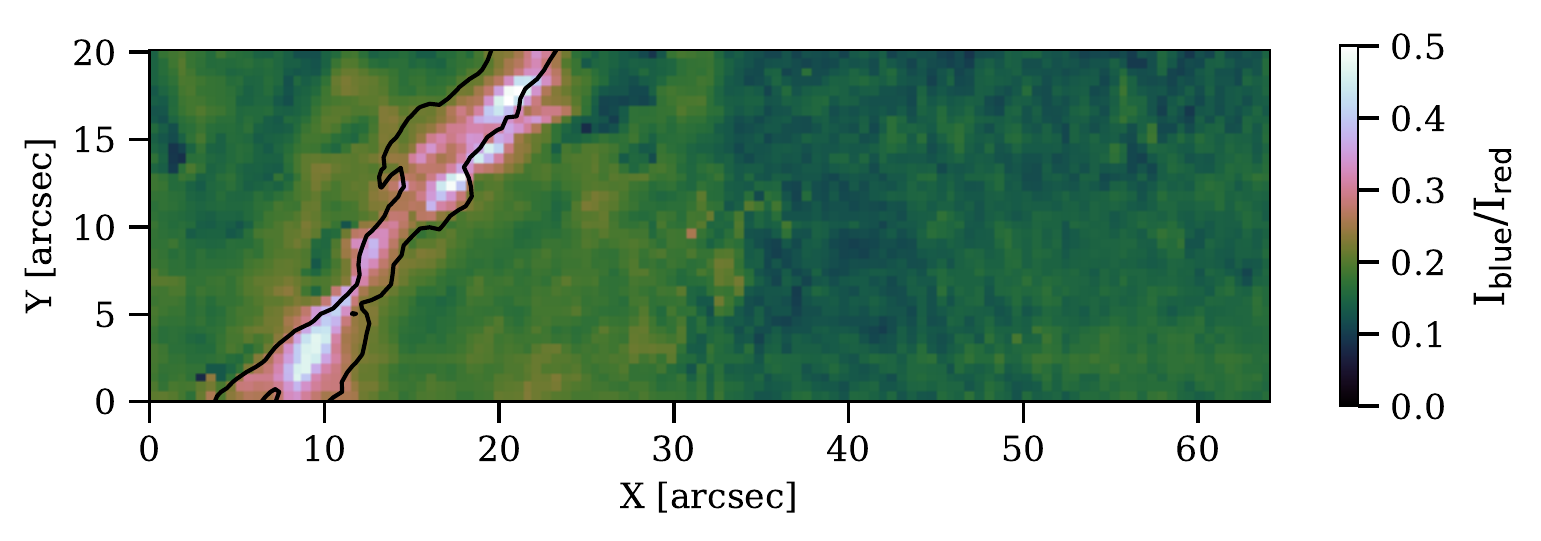}
\caption{Ratio between the Stokes $I$ amplitudes of the blue $A_{B}$ and the red  $A_{R}$  component. The contour shows the level 0.4$A_{R}$,  where the red component starts to saturate.}
\label{fig:ratioArea}
\end{figure}

Figure~\ref{fig:ratioArea} shows the intensity amplitude ratio map between the blue and the red components of the \ion{He}{i} 10830\,\AA\ triplet. Given that the lines are in absorption, we have modeled each component with a Gaussian function with the same width. In those cases in which a second absorption profile appears shifted in wavelength, we have fitted two sets of Gaussians to remove this new component outside the calculations. As the width is roughly the same for the two components of the multiplet, the ratio of amplitudes ${\cal R}=A_{B}$/$A_{R}$ is also approximately equal to the ratio of areas. The line ratio between the two resolved \ion{He}{i} components is around 0.4 inside the filament, whereas it is in the range 0.1--0.2 in the remaining areas \citep{penn95}, the value expected for an optically thin plasma. This means that we find some indication of saturation only in the filament, with a maximum  value of ${\cal R}=$0.5.

Another feature that suggests the presence of  radiative transfer effects is that \Hazel\ systematically produces misfits in the blue component of the multiplet in Stokes $I$. This indicates that the source function used in \Hazel\ that comes directly from the lower atmosphere boundary condition is not appropriate. For this reason, we have added a new multiplicative parameter $\beta$ in \Hazel\ that arbitrarily modifies the source function of the slab $\mathbf{S}$. Consequently, the emergent Stokes profiles are now given by \citep[see][]{asensio2008}:
\begin{equation}
  \mathbf{I} = e^{-\mathbf{K}^* \tau_R} \mathbf{I}_\mathrm{sun} + (\mathbf{K}^*)^{-1}
  \left( 1 - e^{-\mathbf{K}^* \tau_R} \right) \beta \mathbf{S} \ .
\end{equation}
For $\beta<1$ we can mimic a slab whose source function is much smaller than  that determined by the boundary condition. The opposite happens when $\beta>1$, which can also be used to mimic strong emission profiles, like those of flares \citep{Tine2018}. {In this equation $\tau_R$ represents the optical depth measured at the central wavelength of the red blended component of the \ion{He}{i} triplet.}

\begin{figure}[!ht]
\centering
\includegraphics[width=0.44\linewidth]{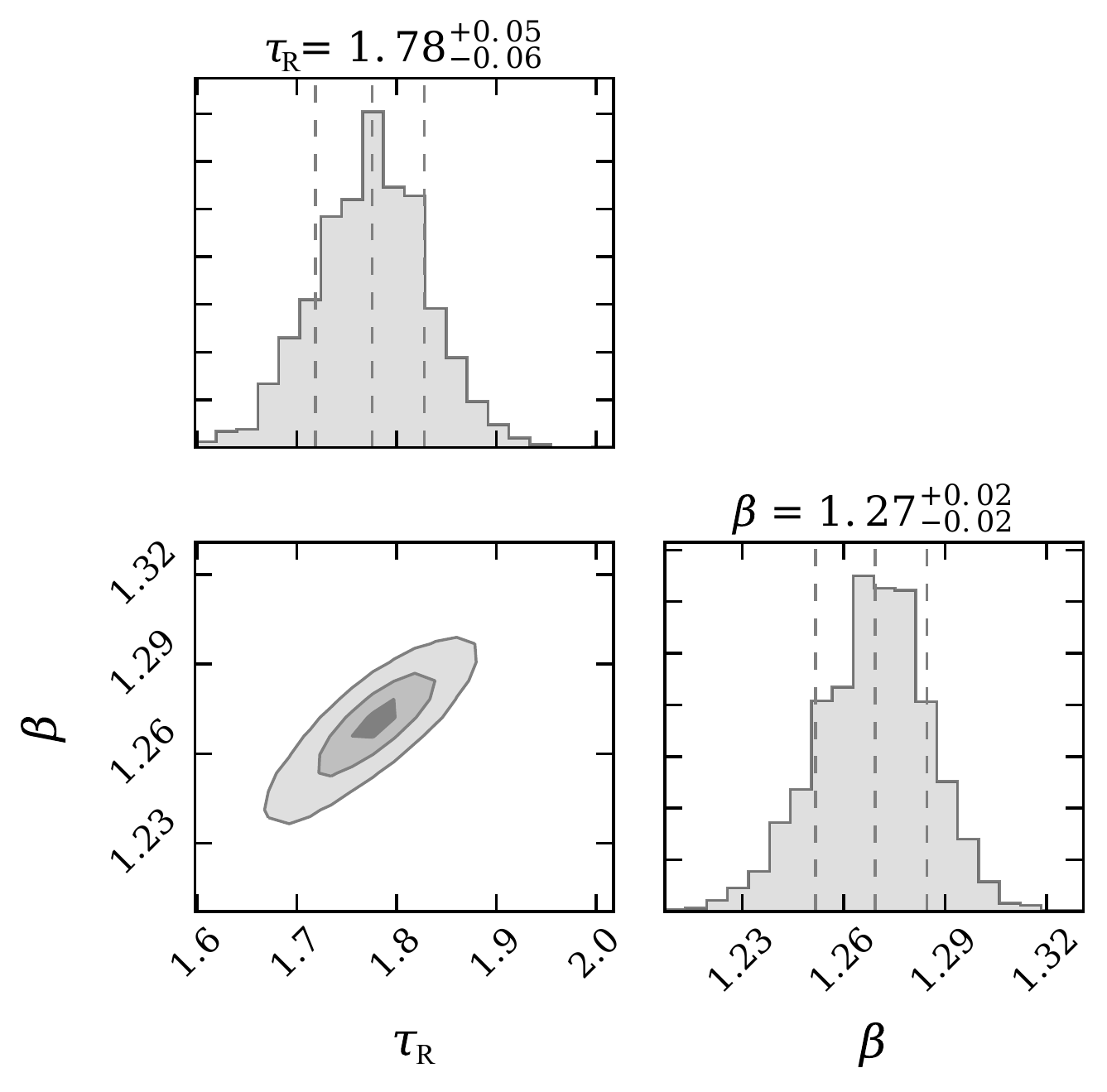}
\includegraphics[width=0.55\linewidth]{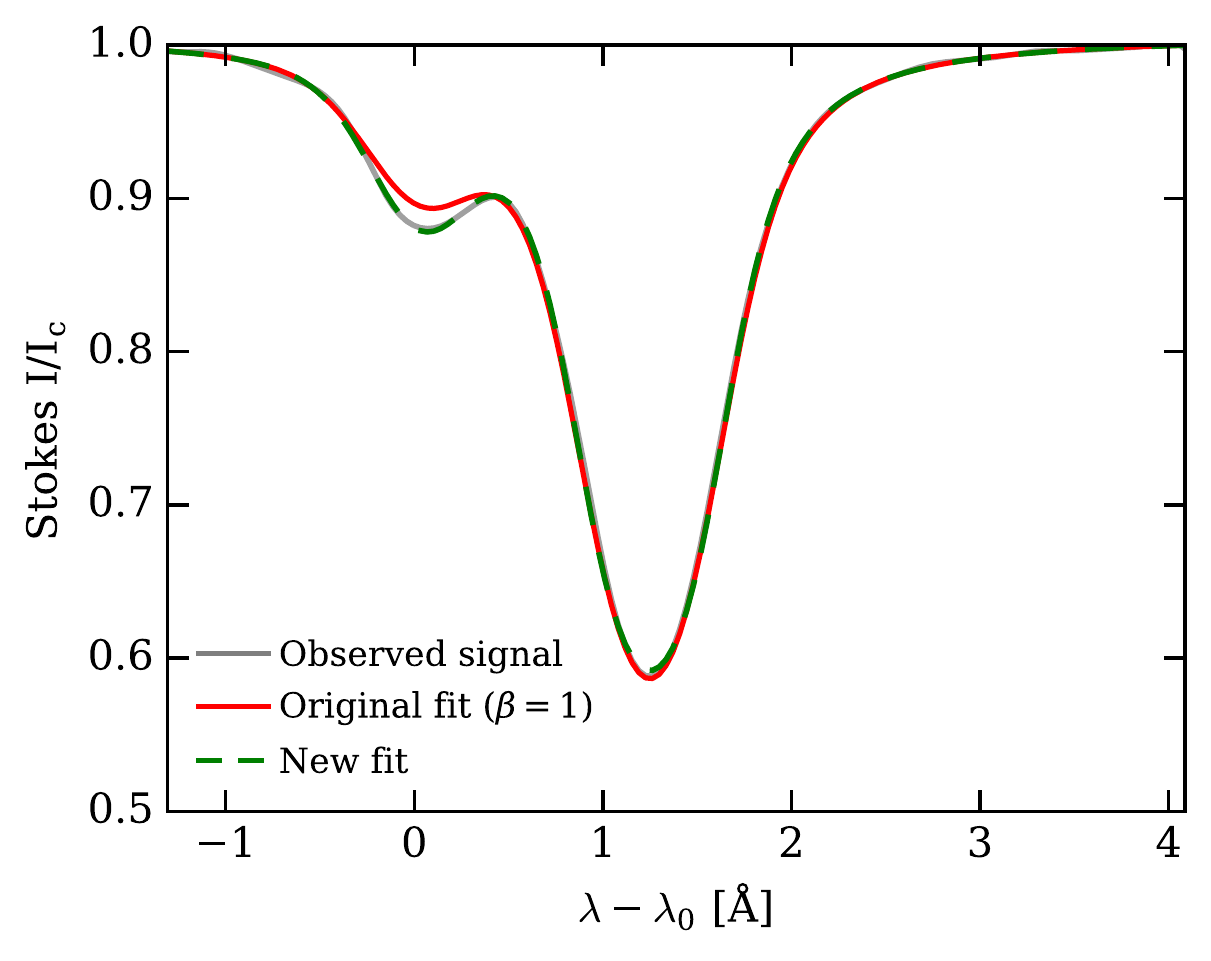}
\caption{Bayesian estimation of the $\beta$ parameter. The left panel shows the posterior probability distribution of $\beta$ and $\tau_R$, and the correlation between them. The right panel shows the improved fit of the blue component of the \ion{He}{i} triplet when going from $\tau_R=1.2,\beta=1.0$ (red line) to $\tau_R=1.8,\beta=1.3$ (green line).}
\label{fig:beta}
\end{figure}

All inversions up to now have been carried out with $\beta$=1. We explore here the effect of adding $\beta$ in the inversion as an additional free parameter. We show in Fig.~\ref{fig:beta} an example of such analysis carried out with \MultiNest\ \citep{feroz2009} to detect possible ambiguities. It is clear that the addition of $\beta$ results in a much better fit of the blue component of the multiplet (green line in the right panel of Fig.~\ref{fig:beta}). However, $\beta$ and $\tau_R$ are degenerate and have opposite impacts in the emergent profile: increasing $\beta$ reduces the absorption, while increasing $\tau_R$ increases the absorption. Therefore, one can compensate (only up to a point) an increase in $\beta$ with an increase in $\tau_R$. In spite of this degeneracy, it is possible to find a pair of $\beta$ and $\tau_R$ that produce a good fit to the Stokes $I$ profile of the multiplet. The general trend is that adding $\beta$ as a free parameter to produce a better fit of Stokes $I$ leads to a slight increase in $\beta$ with a larger increase in $\tau_R$. Figure~\ref{fig:beta} shows this clear correlation between them.

Therefore, from this experiment, we see that the source function needed to reproduce the profiles is greater than in the case of pure scattering ($S=J^0_0$ and $\beta=1$). One possible explanation of this increase might be that the large density of the filament leads to a larger contribution of thermal processes. This idea seems feasible (also suggested by \citealt{CASINI2009}) since our filament has a high absorption (optical depths up to 2 in several scans). Since \Hazel\ neglects collisions in the calculation of the source function (and consequently the thermal contribution), it can lead to a reduction in the polarization signals. Therefore, the magnetic inference in the filament can be affected by this approximation.

\section{Analysis of Stokes $U$}\label{subsec:10830Sign}

\begin{figure}[!ht]
\centering
\includegraphics[width=\linewidth]{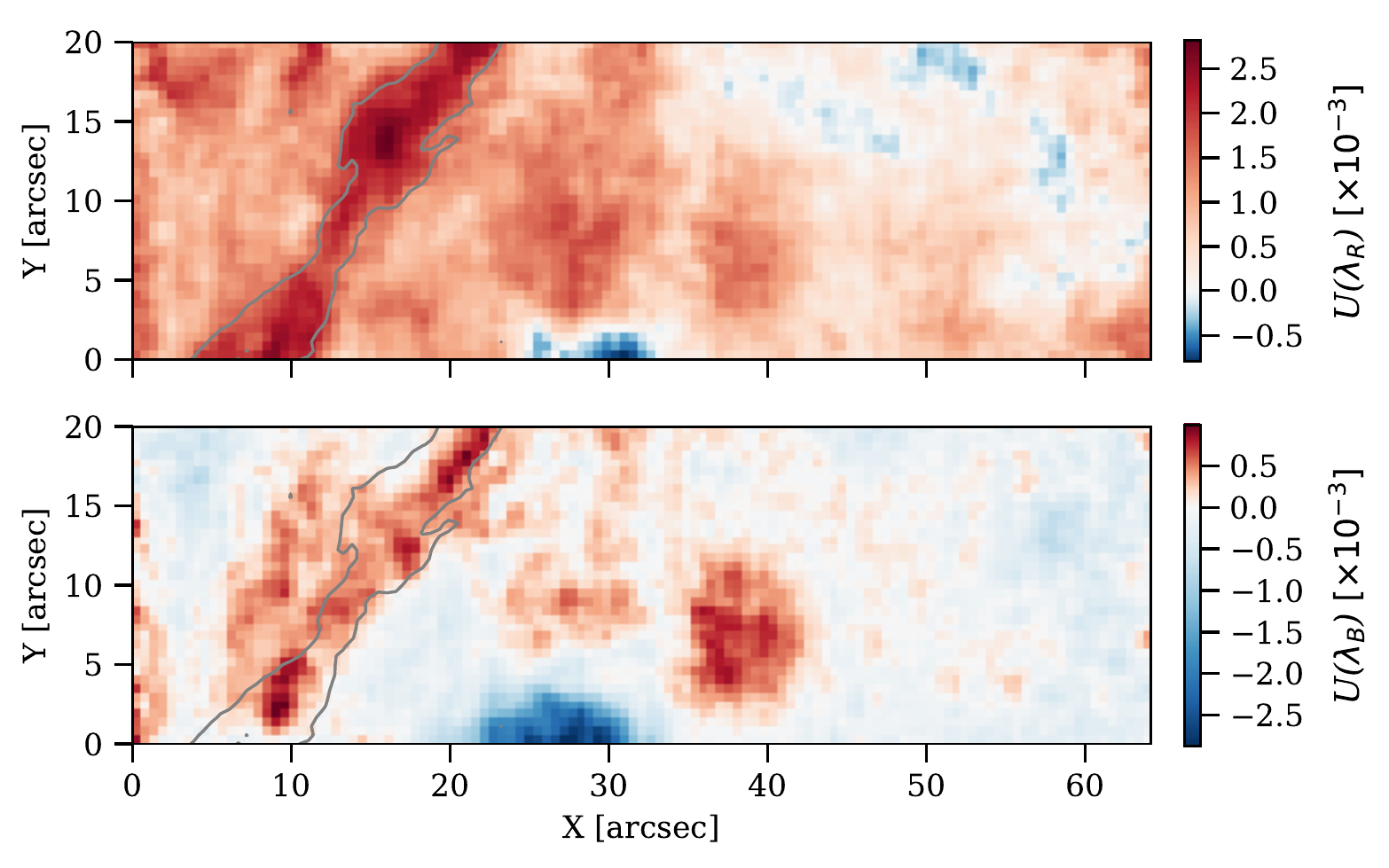}
\caption{Stokes $U(\lambda_R)$ and $U(\lambda_B)$ of the third scan. A grey contour at 0.6$I_{C}$ shows the position of the filament.}
\label{fig:LinearPolRatio}
\end{figure}

As mentioned in Paper~I, the observations display uncommon signals in  Stokes $Q$ and $U$ with the same sign in both the blue and red components. In this section, we perform a detailed study of such signals. Figure \ref{fig:LinearPolRatio} displays maps of the Stokes parameter $U$ (as the polarization signals are larger than in Stokes $Q$) at two different wavelengths: the central wavelength of the blue component, $\lambda_B$ (lower panel), and the central wavelength of the red component, $\lambda_R$ (upper panel). We focus on Stokes $U$ in the third scan because the signals are larger. Similar results are found for the other scans.

We have carefully checked that the observed signals are not produced by any residual crosstalk from Stokes $I$ to Stokes $Q$ and $U$. Given that the red component is several times stronger than the blue one in Stokes $I$, we should have noticed any possible crosstalk first in the red component, and it would have been even more obvious in the silicon line.

The signals outside the filament are generated in a regime in which atomic polarization and the Zeeman effect operate simultaneously. This can be seen in the penumbra of the sunspot, where a very strong and horizontal magnetic field is present. There, two patches are easily observed in the blue component (with opposite signs), but no so easily in the red one. The explanation of this lies on the fact that the atomic polarization affects more the red component than the blue one.  Moreover, the most striking feature of Fig.~\ref{fig:LinearPolRatio} is that both maps present the same sign inside the filament, and opposite sign outside. In the following, we address this issue and try to find a suitable physical scenario that can produce linear polarization signals with the same sign in the two components. In principle, we point out that the only way to generate both signals with the same sign with the current \Hazel\ is by specific magnetic field configurations\footnote{We caution the reader that there may be other options outside the limitations of the forward modeling of \Hazel\ to generate the same effect.}.

\begin{figure}[!ht]
\centering
\includegraphics[width=0.49\textwidth]{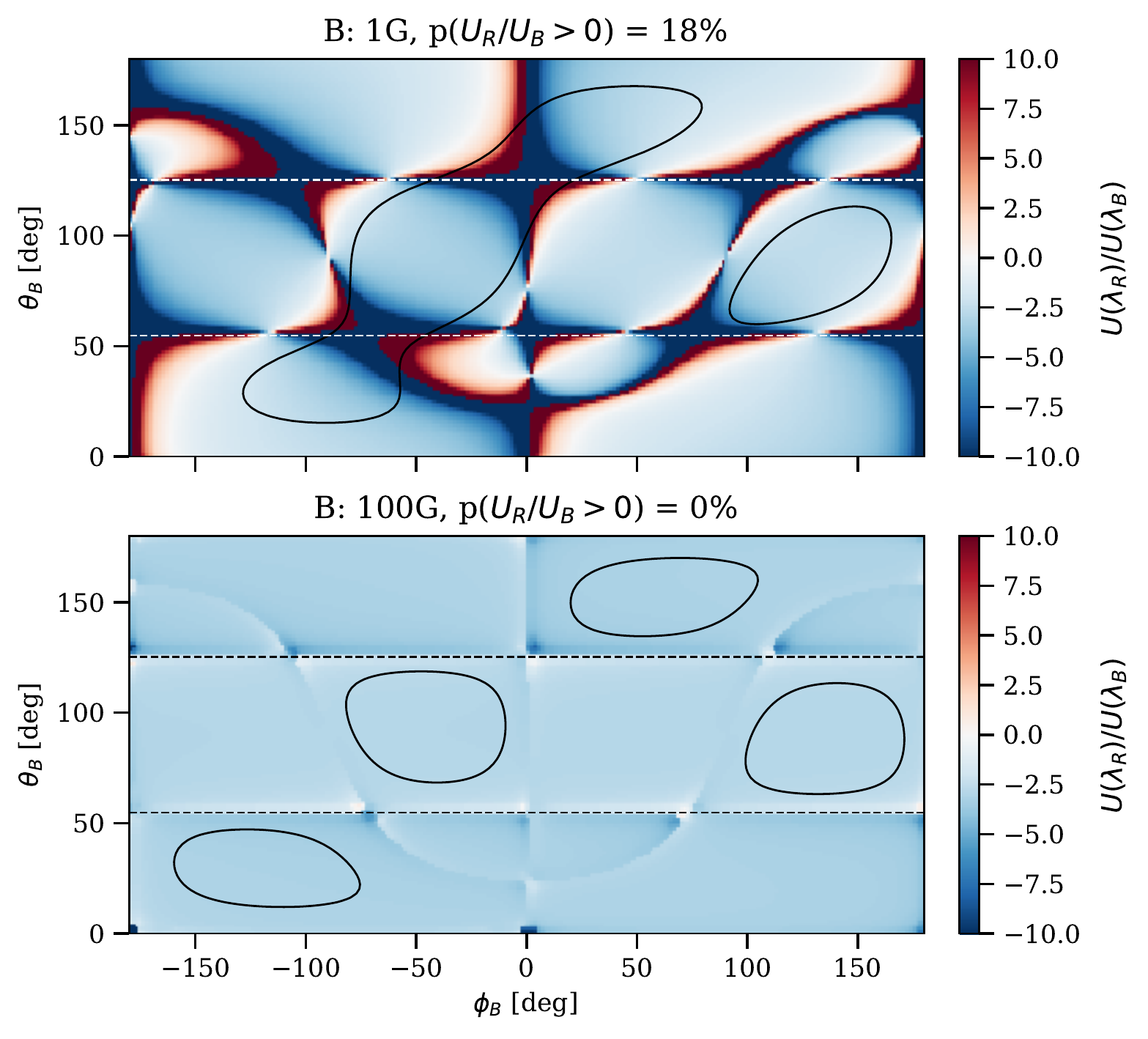}
\caption{Ratio between the synthetic Stokes $U$ signals at the red component $(\lambda_R)$ and the blue component $(\lambda_B)$ for a magnetic field strength of 1\,G (upper panel) and 100\,G (lower panel). Horizontal dashed lines indicate the two Van Vleck angles at $\theta_B=$54.74$^\circ$ and $\theta_B=$125.26$^\circ$. The label on top of each panel describes the percentage of area covered by a positive ratio. Contours indicate the areas where Stokes $U(\lambda_R)>1.5\times10^{-3}I_c$.
}
\label{fig:SyntheticMaps}
\end{figure}

The lower-level Hanle effect (so called because it is produced by the populations and coherences of the lower level) can potentially create linear polarization with the same sign in the blue and red components of the \ion{He}{i}~10830\,\AA. Figure~\ref{fig:SyntheticMaps} shows the ratio $U(\lambda_R)/U(\lambda_B)$ for two different magnetic field strengths and all possible orientations of the magnetic field vector parametrized with the inclination $\theta_B$ and the azimuth $\phi_B$ in the local reference frame. The rest of the parameters are the typical values obtained in the inversion, i.e., $\tau_R=1$, $\Delta v_D=10$\,\kms, $v_\mathrm{LOS}=0$\,\kms, and $a=0.1$. The case of $B=100$\,G (lower panel of Fig.~\ref{fig:SyntheticMaps}) shows that the ratio is relatively constant and around --3 for all configurations of the magnetic field vector. For this value of the magnetic field strength, we see that the probability of finding the two components with the same sign is very small. On the contrary, when the field is decreased to $B=1$\,G (upper panel of Fig.~\ref{fig:SyntheticMaps}), we find that 18\% of the space of parameters gives profiles with the same sign.

These conditions would support the idea of filaments with very weak fields in which  the lower-level Hanle effect produces linear polarization of the same sign for the red and blue components. However, the amplitudes of the linear polarization signals produced for such weak fields are very small, if compared with the observed typical $U(\lambda_R)$ signals of around $1.5\times 10^{-3}I_c$ (marked with contours in Fig.~\ref{fig:SyntheticMaps}). Additionally in such cases the synthetic Stokes $Q$ profiles have larger amplitudes than the observed signals. Therefore, we conclude that the linear polarization profiles cannot be explained (only) with very weak magnetic fields, with a single component slab model and under the assumptions of the version of \Hazel\ at the time of writing. There are other plausible explanations which will be discussed in the conclusions. 

\section{Analysis of Stokes $V$}\label{sec:stokesv}

In Paper~I we showed that the circular polarization maps in the chromosphere  displayed a pattern very similar to the photospheric map without any indication of the presence of the filament. In addition, in \cite{diaz2016} we demonstrated that active region filaments, despite being dense chromospheric structures embedded in the corona, do not have enough optical thickness so as to completely block the light emerging from the underlying chromosphere. At this point it is not clear whether the filament has a magnetic field configuration that resembles the photospheric magnetic field or, on the contrary, these signals were coming from the underlying chromosphere. In this section, we investigate the presence of more than one magnetic component in the filament region from the observed circular polarization profiles and, in particular, from the amplitude ratio of the red and the blue components of the multiplet. If both components are formed under similar physical conditions, the circular polarization profiles will have a constant amplitude ratio between the two  components. If this ratio has a different value, they can potentially sense different physical conditions given their different optical depths.

\begin{figure}[!ht]
\centering
\includegraphics[width=\linewidth]{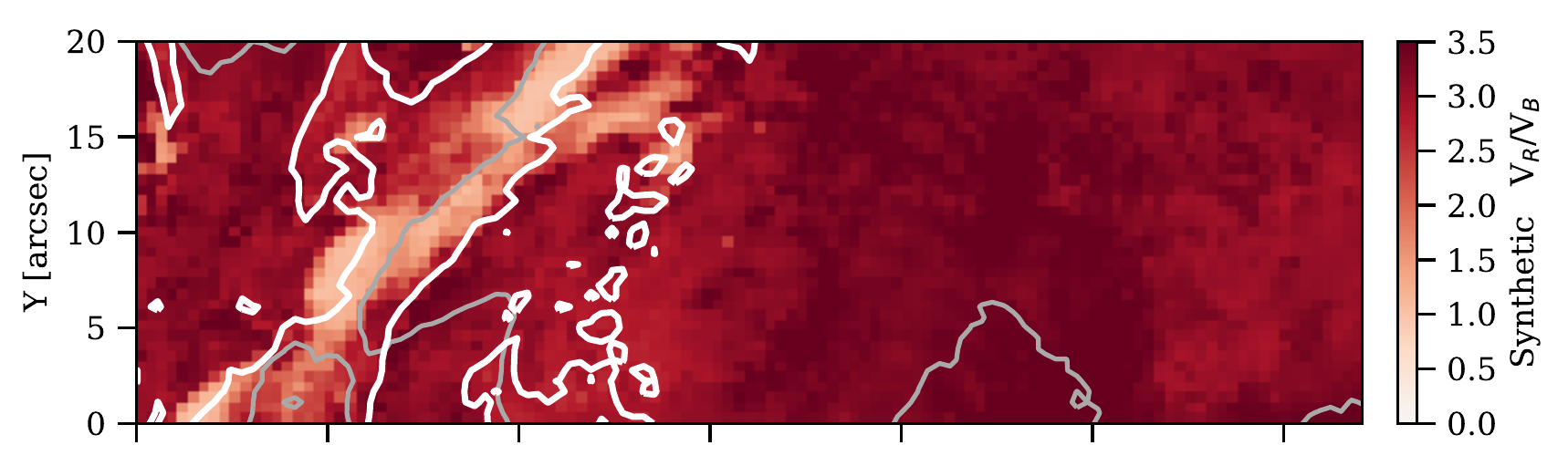}
\includegraphics[width=\linewidth]{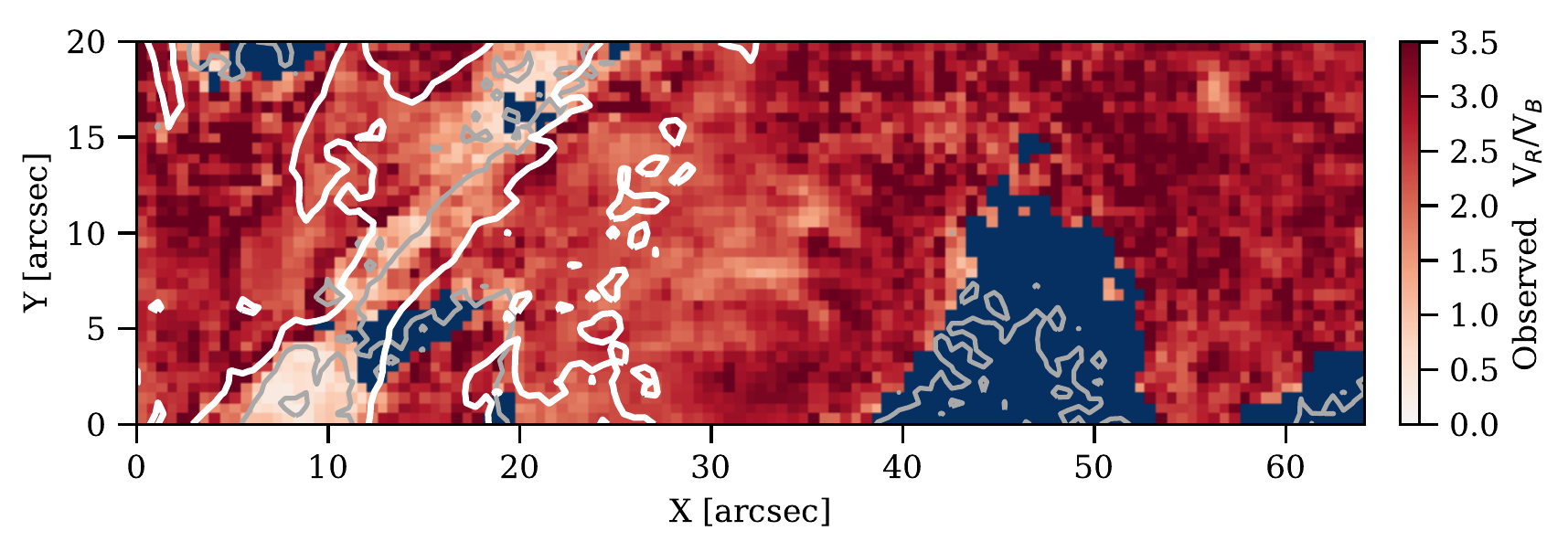}
\caption{Upper panel: map of the ratio of the amplitudes of the circular polarization between the red and the blue components calculated using the synthetic profiles from the \Hazel\ fit. Lower panel: the same ratio but calculated from the observed signals, masking the signals below 1.5$\sigma$. The gray line indicates the PIL while the white contour indicates the filament location.}
\label{fig:ratioV}
\end{figure}

To this end, we compute the ratio of the Stokes $V$ amplitudes calculated in the blue lobe\footnote{We have chosen  the blue lobe because the red lobe of the red component is sometimes affected by a highly redshifted component.} of the red and blue components from the observations ($V_R/V_B$) which is displayed in the lower panel of Fig.~\ref{fig:ratioV}. To compare with the single component model inferred by \Hazel, we show the same ratio for the synthetic profiles in the top panel of Fig.~\ref{fig:ratioV}. The observed ratios shows very small values in the polarity inversion line (gray line in the figure), which are produced by the presence of noise. To avoid artifacts produced by the noise, we mask all values with signals below a certain threshold. To set this threshold we find that the estimation of the ratio is only reliable when the Stokes $V$ signals are above 1.5 times the standard deviation of the noise.

\begin{figure}[!ht]
\centering
\includegraphics[width=\linewidth]{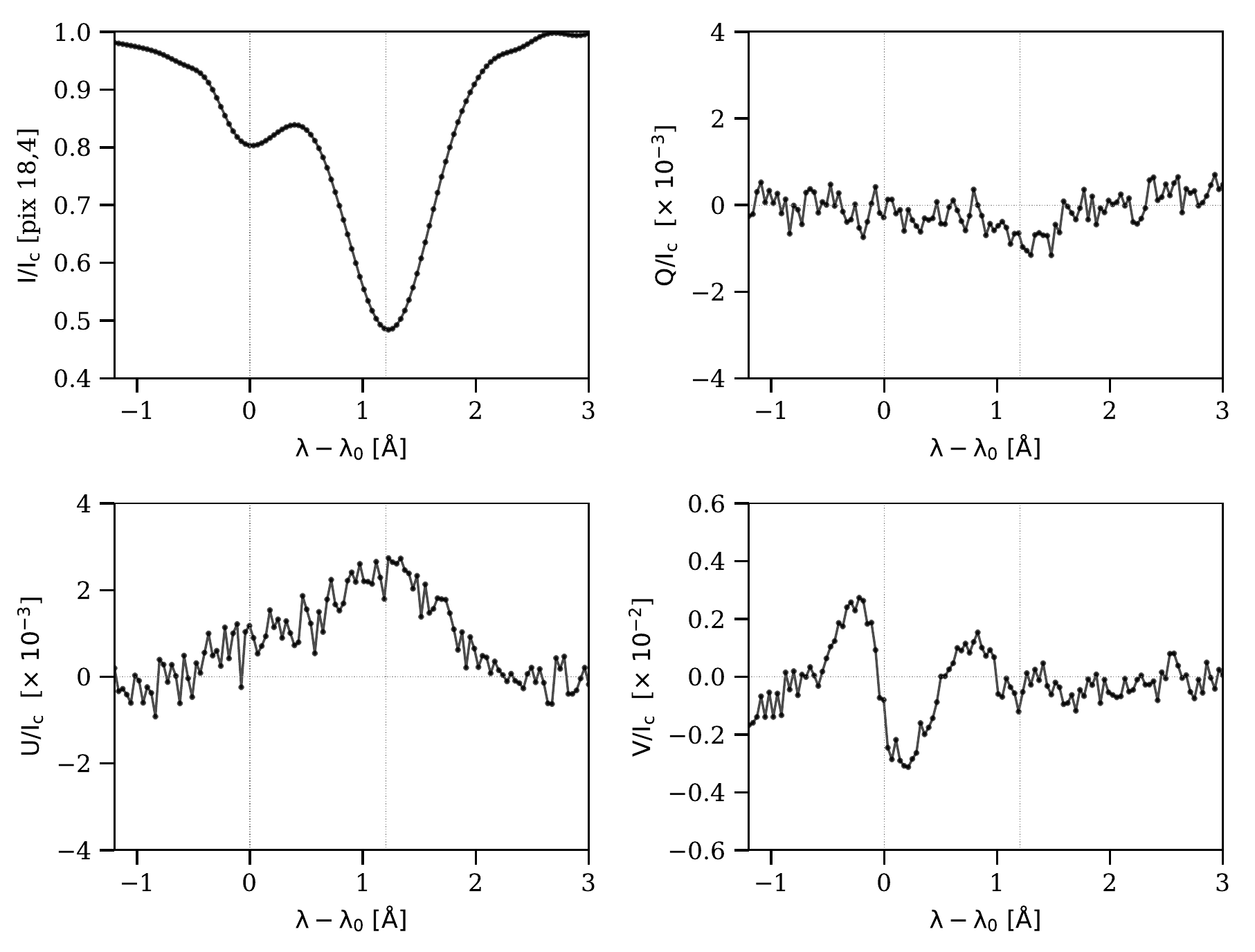}
\caption{Stokes profiles of the pixel [18,4] with very low ratio $V_R/V_B$, located in the lower part of Fig.~\ref{fig:ratioV}. The Stokes $V$ amplitude of the blue component of Stokes $V$ is twice the one of the red component and $\lambda_0$ is the wavelength of the center of the blue component.}
\label{fig:sinred}
\end{figure}

The similarities between the two maps in Fig.~\ref{fig:ratioV} indicate that the code is able to reproduce the observed  asymmetries between the components of the line. A large fraction of the FOV shows observed and synthetic ratios around 3 except inside the filament, where we find lower values (around 1.0). The decrease of this ratio in the filament is mainly a consequence of the saturation of the red component due to the high optical depth. A single component model seems to be sufficient though to reproduce many of the filament profiles. However, we find pixels in the FOV with ratios much smaller than those that can be produced by the saturation of the spectral line. The limit for a single component atmosphere is 1, but values as low as 0.2 are found in the observations. This indicates that more complex models with magnetic field gradients have to be used to explain such big differences. An example of this case is displayed in Fig.~\ref{fig:sinred} which has been extracted from the region at (8\,\arcsec, 3\,\arcsec). In this figure the Stokes $V$ amplitude of the blue component is almost twice that of the red component.

The question that now arises is whether the rest of the profiles, even if they do not show such clear evidence in the ratio $V_R/V_B$, can be explained as profiles emerging from a multi-component model. To visualize this and to demonstrate that the saturation effect can be produced by either a single-component or a multi-component model, we show in the following how, under particular conditions, a single-component model is able to roughly reproduce the emerging profiles synthesized with a two-component atmosphere (one slab on top of the other as the model described in \citealt{diaz2016}).

For this numerical experiment, we synthesize the emergent Stokes profiles from a simulated filament (top slab) with $\tau_R=1.2$ placed above an active region at the same heliocentric angle as the observations ($\mu=0.92$) with a background absorption (bottom slab) of $\tau_R=0.3$ (average value extracted from our observations). The magnetic field in the active region is parallel to the surface ($\theta_B=90^\circ, \phi_B=0^\circ$) and perpendicular to the filament axis (simulating a polarity inversion line) with a strength of $B=300$\,G. The filament has a magnetic field strength of $B=10$\,G, with an azimuth of $45^\circ$ with respect to the  field underneath ($\theta_B=90^\circ, \phi_B=-45^\circ$). This configuration mimics the general trend of what we observe in the Stokes profiles.

\begin{figure}[!ht]
\centering
\includegraphics[width=\linewidth]{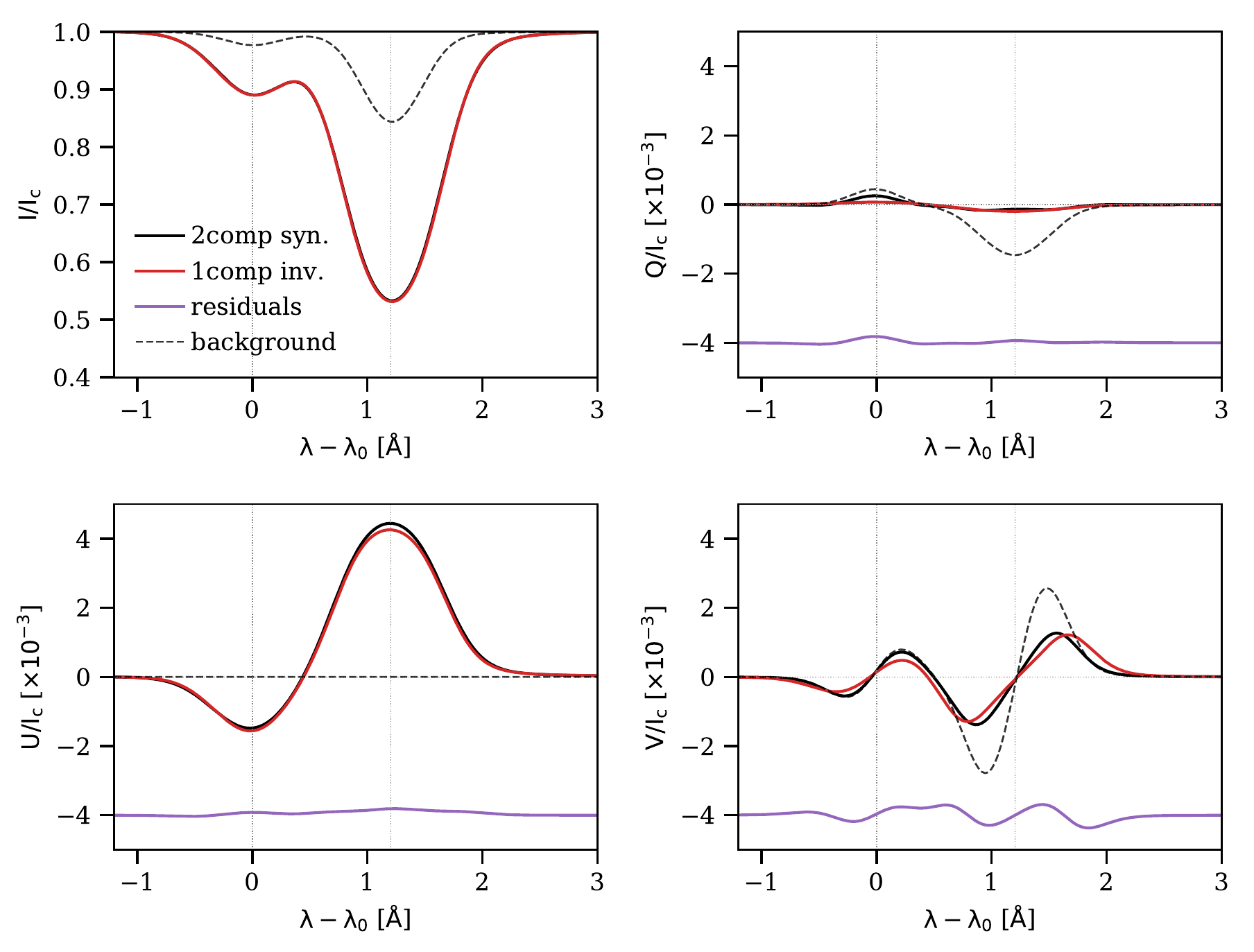}
\caption{One component model inversion fit (red) of the synthetic Stokes profiles generated with a two component model (black). The residuals of the best-fit are below each Stokes profile (purple). The dashed line indicates the Stokes signals from the background in absence of the filament.}
\label{fig:1desde2}
\end{figure}

The synthetic Stokes profiles are displayed in Fig.~\ref{fig:1desde2} together with the profiles of the fit using a single component model. The following conclusions can be extracted from this experiment. i) The optical depth is roughly equal to the sum of those of the two components $\tau_R\simeq 1.5 = 1.2+0.3$. ii) Although the optical depth of the filament is large, the emergent Stokes $V$ from the slab below remains almost unperturbed for the blue component after passing through the second slab, while its amplitude decreases for the red components (the original Stokes $V$ of the background is drawn in dashed line). Some ``transparency'' of the plasma to circular polarization can be expected as the opacity decreases towards the wings precisely where the Stokes V signals are higher, while it is almost zero in the core of the line where the absorption is very high. In the blue component of the \ion{He}{i} 10830\,\AA\ multiplet, because of its lower optical depth, the change is even smaller\footnote{
This can be easily visualized assuming a Gaussian profile for the absorption with width
$\sigma^2$. Then, the Stokes $V$ generated in the first slab will be attenuated by
$V_2(\lambda_0)=V_1(\lambda_0) e^{-\tau_2(\lambda_0)}$ in the core. At the wavelengths on
the Stokes $V$ lobes ($\lambda = \lambda_0 \pm \sigma$) the absorption is $\tau_2 (\lambda_0 \pm
\sigma) = \tau_2(\lambda_0) e^{-0.5}\sim0.6\tau_2(\lambda_0)$. If, for example,
$\tau_2(\lambda_0)=1.2$, then $V_2(\lambda_0\pm \sigma)=V_1(\lambda_0\pm
\sigma) e^{-0.6\tau_2(\lambda_0)}\sim0.5V_1(\lambda_0\pm
\sigma)$, that has only reduced to a half. In the blue component, because of its lower optical depth, the change is less important.
}. iii) The linear polarization detected in Stokes~$U$ is mainly due to the contribution of the filament (top slab) proportional to the absorption as we detected in the polarization maps of Paper~I. The Stokes $Q$ signal generated by the background is attenuated, leaving the weak signal generated by the filament. iv) A single component inversion is able to reproduce the Stokes signals at a precision level of $5-10\times 10^{-5}I_c$, including the high asymmetry between the Stokes~$V$ of the components. The residuals are well below our noise level of $4-6\times10^{-4}I_c$ so that it renders these differences impossible to be detected. iv) From the inversion we have inferred a magnetic field with $B=250$\,G, $\theta_B=78^\circ$, and $\phi_B=-45^\circ$ for the filament, which is much stronger and more vertical than the magnetic field in the upper slab (10\,G), a fact that has to be kept in mind in the analysis of these structures \citep{diaz2016}.

\section{Abrupt changes in the magnetic field}
\label{sec:inner}

\begin{figure}[!ht]
\centering
\includegraphics[width=\linewidth]{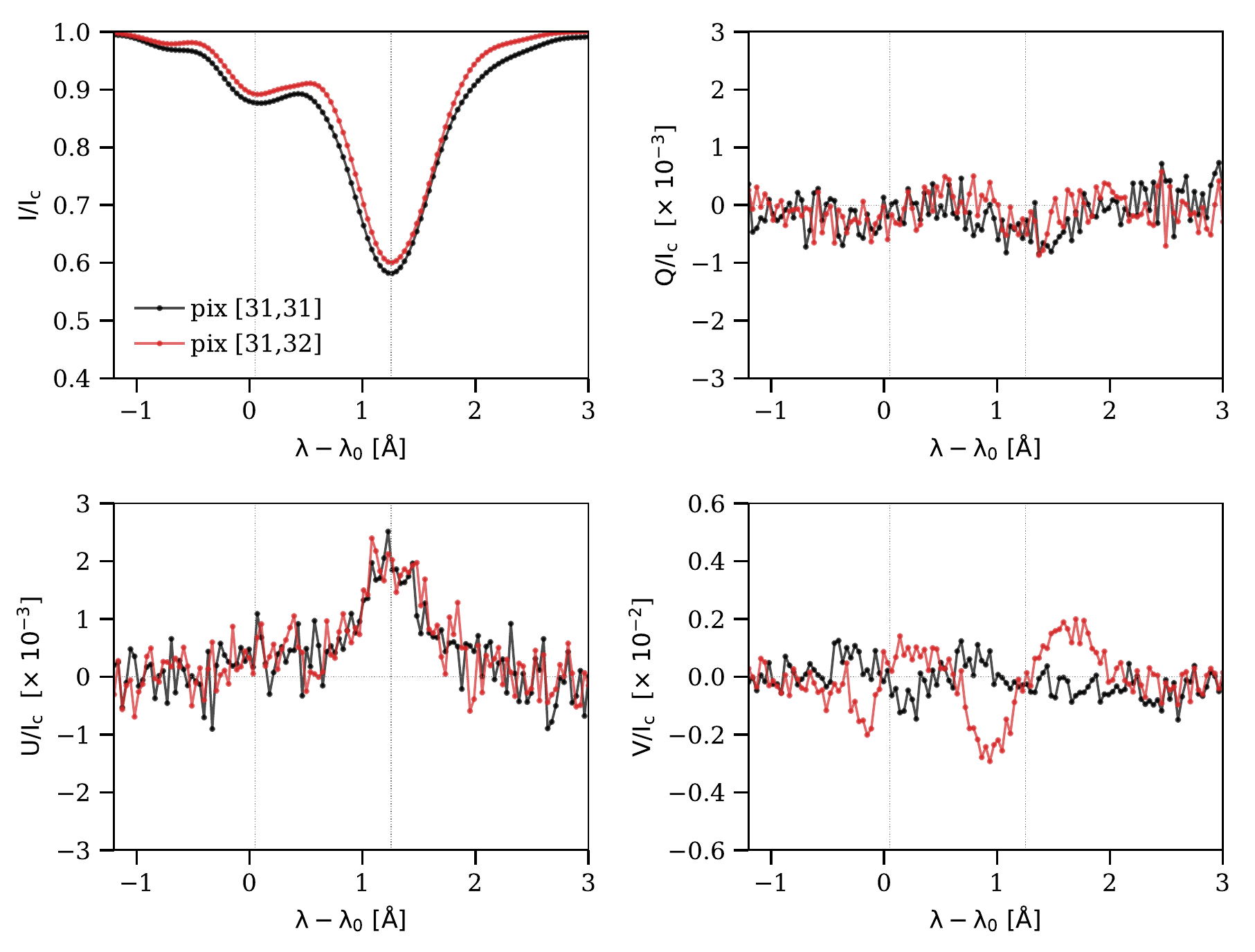}
\caption{Stokes profiles of the two chosen pixels from the inner part of the filament. They are two contiguous pixels at each side of the PIL. Stokes $Q$ and $U$ are very similar, while each Stokes $V$ have a different polarity.}
\label{fig:pixelInside}
\end{figure}

In Paper~I we suggested that the local azimuth map obtained from the one component inversion displayed solutions correlated with the polarity of the region, i.e., the solutions were conditioned by the Stokes $V$ information. Moreover, in the PIL, this polarity change produced an azimuth change of up to 180$^\circ$ within the filament.  Therefore, in order to understand how the topology of the magnetic field really is, it is necessary to study how ambiguous solutions are distributed. In case one is able to discard some of the solutions because of their spatial distribution, the possible configuration of the filament could be obtained. In this section we analyze two contiguous pixels at each side of the PIL (i.e. from the filament axis) only taking into account Stokes $Q$ and $U$ whose signals, as we have seen from the previous analysis, show the properties of the filament. If this abrupt change in the azimuth is real, we expect to retrieve it without the information of Stokes $V$.  The Stokes profiles in these pixels are displayed in Fig.~\ref{fig:pixelInside}.

\begin{figure}[!ht]
\centering
\includegraphics[width=\linewidth]{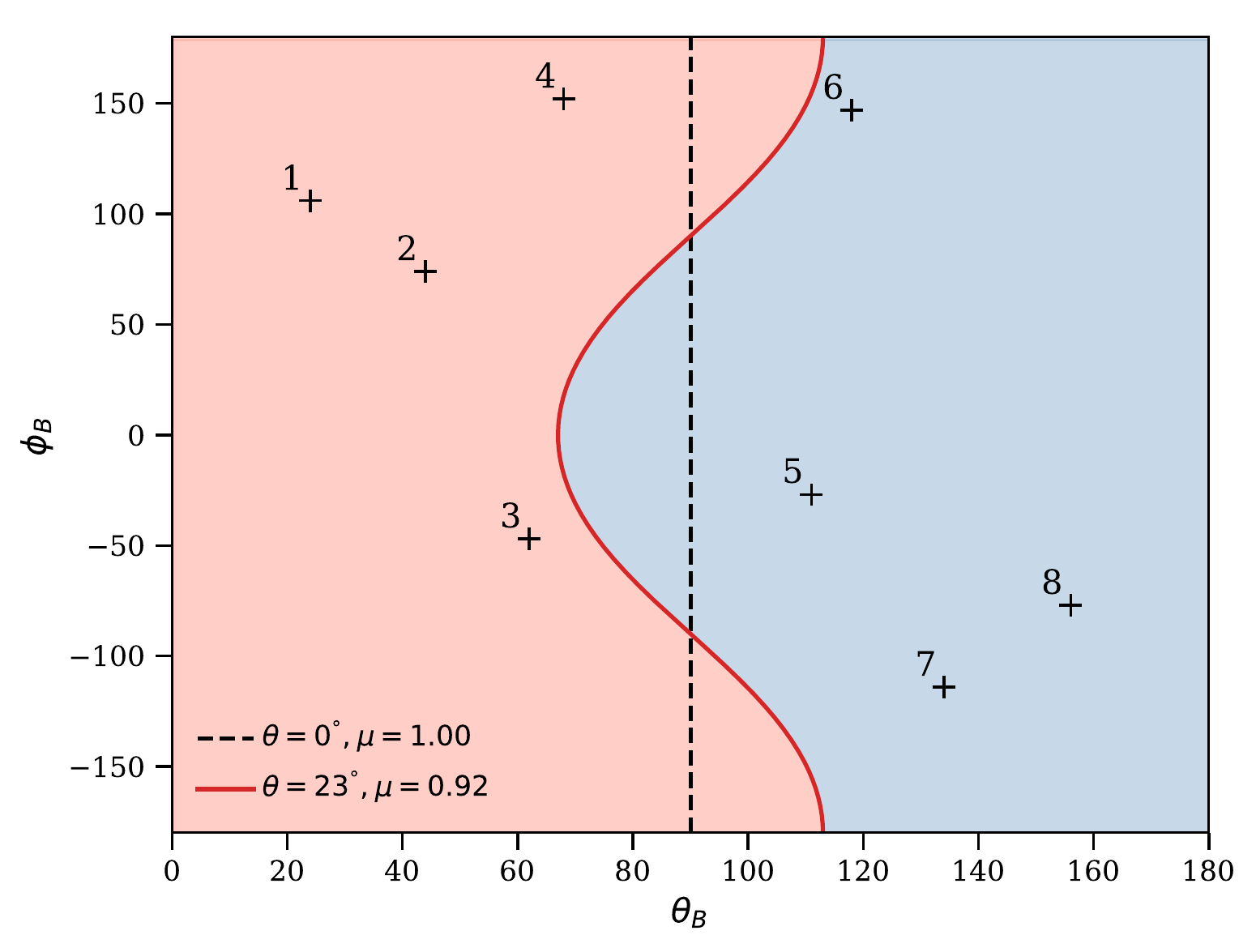}
\caption{Hanle ambiguities represented with plus symbols, in the space of $\theta_B$ and $\phi_B$ compatible with both Stokes $Q$ and $U$ of the two chosen pixels of Fig.~\ref{fig:pixelInside}. The coloured areas correspond to each polarity of the magnetic field, indicating with the dashed (which
degenerates into a single vertical line) and continuous contours the location of the polarity inversion for $\theta=0^\circ$ and $\theta=23^\circ$, respectively.}
\label{fig:Discrepancies}
\end{figure}

Both pixels present very similar linear polarization signals but different polarities in Stokes $V$. As in previous sections, we have used \MultiNest\ to find the locations of the eight potential ambiguous solutions, which are displayed in Fig.~\ref{fig:Discrepancies}. These eight combinations of $\theta_B,\phi_B$ generate the same Stokes $Q$ and $U$ profiles. Figure~\ref{fig:Discrepancies} is split in two regions of different polarities: the red one with the magnetic vector pointing towards the observer and the blue one in the opposite direction. The red solid line shows the location of $\Theta_B=90^\circ$ when it is transformed to the local reference frame $\theta=23^\circ$. This division also splits the eight solutions in two groups of four depending on the polarity of Stokes $V$. This implies that the local azimuth and inclination inferred would be very different only due to the polarity, even for exactly the same Stokes $Q$ and $U$ profiles.

Given that drastic variations of the azimuth of the field between neighboring pixels are discouraged in the filament, one could try to find whether smooth solutions in the full magnetic field vector can be found\footnote{An alternative explanation for this situation is that we are detecting the two branches of a twisted flux rope at different heights but spatially nearby due to the LOS perspective. This case seems unlikely in the center of the filament (the location where these pixels are extracted) because it looks compact/homogeneous, so that the assumption a single plasma slab is plausible. The edge of the filament, where the opacity changes rapidly and where we clearly see fine structure, could be affected by this effect.}. Let us focus on the pair of solutions closest in local azimuth and with different polarities, the pair [4, 6] (Fig.~\ref{fig:Discrepancies}). In this case, the two solutions differ by  $\Delta\theta_B=120^\circ-70^\circ=50^\circ$ in inclination. {Consequently, to simultaneously fit a similar Stokes $V$ signal of opposite polarity, one needs to modify the magnetic field strength from $B=20$\,G$,\theta_B=70^\circ$ to $B=200$\,G$,\theta_B=120^\circ$. This increase in magnetic field strength is so large because solution 6 has a magnetic inclination in the plane of the sky close to $\Theta_B=90^\circ$} (near the red line in Fig.~\ref{fig:Discrepancies}). In conclusion, ensuring smoothness in the azimuth does not ensure smoothness in the inclination and/or the magnetic field strength. This is another indication that these Stokes profiles cannot be analyzed assuming a single component model and we need more than one component to interpret them.

\section{A two component inversion}

After finding several evidences of the necessity of more complex models to explain the observed signals, in this section we explore the possibility of a two component inversion with the idea to separate the contribution of the filament from the region below. This scenario is based on the assumption of two slabs of constant physical properties, one on top of the other, where the emergent polarized light have to pass through the two of them (as in the example of Sec.~\ref{sec:stokesv}). In principle, although each slab can have different velocities, absorptions, and line broadenings, we reduce the dimensionality of the problem by only doing the inference over the magnetic field vector of both components:  $(B_1,\theta_1,\phi_1,B_2,\theta_2,\phi_2)$. The rest of parameters have been fixed to those obtained from a single component inversion.

\begin{figure}[!ht]
\centering
\includegraphics[width=\linewidth]{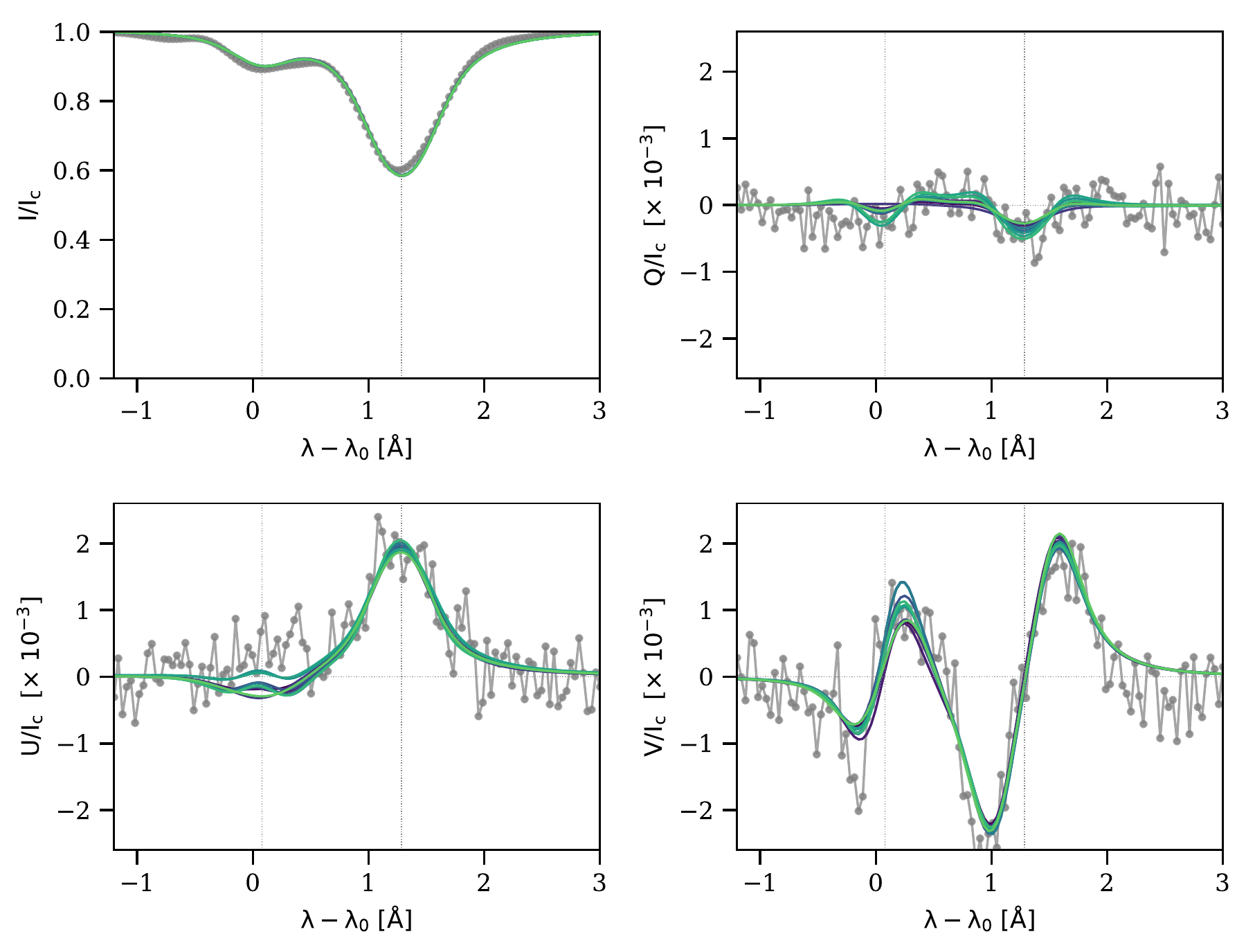}
\caption{From green to blue colors, 10 of the 30 compatible solutions with the observed  profile (grey).}
\label{fig:MC2C}
\end{figure}

Both slabs have the same thermodynamic properties but different optical depths $\tau_1=0.3$ (value estimated from the filament surroundings) and $\tau_2=0.9$ (to mimic our average filament absorption). In this exercise we have chosen the pixel [31, 32], already used in Fig.~\ref{fig:pixelInside}. The observed Stokes profiles and the synthetic ones found for each mode of the $\chi^2$ surface are displayed in Fig.~\ref{fig:MC2C}. We see that it is possible to reproduce almost all observational features, such as the positive Stokes $U$ in the blue component or the complex Stokes $Q$ shape, which is impossible with a single component.

However, the inference problem becomes very complex because we find a high number of compatible solutions, all of them with different configurations. In this case, we have found up to 30 solutions but only 10 are shown to avoid crowding the figure. These solutions are spread in the space of parameters with an average distance between each one around 10$^\circ$ in the inclination and azimuth and 30\,G in the field strength $B$. The high number of solutions is a consequence of the combination of all possible ambiguities in both components and the flexibility of the two component model to generate the same profiles with different combinations of parameters.

Moreover, because of the possible cancellation of the addition of signals when crossing the two components \citep{diaz2016}, the magnetic field can sometimes be increased (if their signals are opposite) almost arbitrarily, finding cases in which magnetic fields of 500--800\,G are required to reproduce the profiles, while a single component inversion yields magnetic fields below 100\,G. Given the multi-parametric nature of this analysis, and after extensive numerical experiments (adding or fixing other parameters) we conclude that a two component inversion with all the free parameters is a challenging problem if there is no extra information to constrain them.

\section{Summary and conclusions}\label{sec:summary}

After a first canonical analysis of an active region filament observed in the \ion{He}{i} 10830\,\AA\ multiplet using a single component model, in this second part of the series we have focused our attention on observational evidences of the necessity of more complex models to explain the observations. As the first evidence, the intensity profiles points to a very optically thick filament where \Hazel\ cannot perfectly reproduce the profiles, underestimating the absorption of the blue component. We fixed it by using a new parameter which takes into account the extra emission of the slab. However, this improvement is not enough if radiative transfer effects make both components of the multiplet be sensitive to different physical conditions.

The second is the ubiquitous presence of profiles with the same sign in the blue and red components in the linear polarization inside the filament. Because a single component cannot reproduce these observational features, we mention the proposals by \cite{AsensioMaser2005,TB2007,judge2015} to explain it. They show that if we allow for radiative transfer effects, the variation of the anisotropy of the radiation field, $J^2_0$, in an optically thick slab along the LOS can lead to the same sign in both component. Due to the high opacity, the red and blue components are sensitive to different regions in the atmosphere with different $J^2_0$ values of opposite signs, resulting in polarization signals with the same sign. This seems to be a plausible scenario for active region filaments because they are located at low heights (where the anisotropy can change its sign more easily than at larger heights) and they have a significant optical thickness (to be sensitive to different regions in each component and promote the horizontal illumination). These two ingredients favor a negative contribution to the anisotropy because the self-emission of the slab can compete with the anisotropy of the photospheric radiation field, producing these changes of sign of $J^2_0$ with depth. Although this possibility seems realistic, the calculations of these competing effects in the previous studies use still simple assumptions. In reality, one should use 
a complex multi-slab model where the variation of anisotropy along the slab and the formation of both components must be explicitly taken into account. It remains to be checked whether this effect can still hold in the complex scenario. Again, this is an additional evidence supporting the necessity of increasing the complexity of the current models.

The Stokes $V$ map of \ion{He}{i} does not show any clear signature of the presence of the filament, and the inferred azimuth map follows the same pattern than Stokes $V$, as if the polarity of Stokes $V$ was conditioning the inference. By using simple numerical experiments we have demonstrated here \citep[see also][]{diaz2016} that the filament is almost transparent to the circular polarization generated in the atmosphere below the filament. Its analysis as a signal coming from a single slab leads to discrepancies, such as the abrupt changes in the azimuth or assigning larger field strengths to the filament. Moreover, some indications of strong gradients of $B$ are also visible in Stokes $V$.

Finally, in an effort to study the reliability of two-component inversions with the idea of separating both contributions, we have shown that the model contains too much flexibility. Although it can reproduce better the observations (better than a single component model),  a large number of compatible solutions with different configurations exist, making the interpretation very complex. To overcome this problem and enable the analysis of the filaments, we propose stereoscopic observations (to identify the height) and magnetic field extrapolations from the photosphere or multi-line observations with for instance the \ion{Ca}{ii} at 8542\,\AA\ \citep{Khomenko2016,Schwartz2016} to infer the magnetic field of the active chromosphere below. For these optically thick structures close or above the disk (also spicules, archs, etc.) radiative transfer modeling and extra information about the illumination from the photosphere is also needed. However, other cases where the observed line is optically thin and there is no background (such as in prominence observations using the D$_3$ at 5876\,\AA, \citealt{Casini2009B}), a single component could be a plausible option. Finally, it is also important to add that the presence of noise has a strong impact on the inference process, in which a lower noise level can help reduce the number of ambiguous solutions.

The importance and misinterpretation due to oversimplified models have been also investigated by other studies in the past. An example is \cite{milic2016}, who described the bias produced when a simple 1D model is used to interpret a 2D inhomogeneous prominence model, retrieving more vertical and weaker magnetic field solutions. Our contribution describes to the community the details that we should take into account when interpreting data from active region filaments when the next generation of telescopes arrive. The problem is complex and we should be prepared to deal with all problems to further improve our knowledge of the chromosphere.

\begin{acknowledgements}

The authors wish to thank Manolo Collados for useful advice on the presented work. We specially thank Andreas Lagg for observing the active region filament and providing this valuable data for us. We want also to thank the anonymous referee for the comments and suggestions.

The 1.5-meter GREGOR  solar telescope was built by a German consortium under the leadership of the Kiepenheuer Institut f\"ur Sonnenphysik in Freiburg with the Leibniz Institut f\"ur Astrophysik Potsdam, the Institut f\"ur Astrophysik G\"ottingen, and the Max-Planck Institut f\"ur Sonnensystemforschung in G\"ottingen as partners, and with contributions by the Instituto de Astrof\'sica de Canarias and the Astronomical Institute of the Academy of Sciences of the Czech Republic. The GRIS instrument was developed thanks to the support by the Spanish Ministry of Economy and Competitiveness through the project AYA2010-18029 (Solar Magnetism and Astrophysical Spectropolarimetry).

Financial support by the Spanish Ministry of Economy and Competitiveness through projects AYA2014-60476-P and AYA2014-60833-P are gratefully acknowledged.

CJDB acknowledges Fundaci\'on La Caixa for the financial support received in the form of a PhD contract.

This research has made use of NASA's Astrophysics Data System Bibliographic Services.

This paper made use of the IAC Supercomputing facility HTCondor (\url{http://research.cs.wisc.edu/htcondor/}), partly financed by the Ministry of Economy and Competitiveness with FEDER funds, code IACA13-3E-2493.

We acknowledge the community effort devoted to the development of the following open-source packages that were used in this work: \texttt{numpy} (\texttt{numpy.org}), \texttt{matplotlib} (\texttt{matplotlib.org}), \texttt{corner} \citep{corner}, and \texttt{SunPy} (\texttt{sunpy.org}).

\end{acknowledgements}

\bibliographystyle{mnras}

\end{document}